\documentclass[prc,twocolumn,showpacs,superscriptaddress,floatfix]{revtex4}

\usepackage{graphicx}

\begin{document}               

\def\be{\begin{equation}}
\def\ee{\end{equation}}
\def\ba{\begin{eqnarray}}
\def\ea{\end{eqnarray}}
\def\bas{\begin{eqnarray*}}
\def\eas{\end{eqnarray*}}
\def\bareps{\overline{\varepsilon}}

%\addtolength{\topmargin}{2cm}

\title{Density-functional theory for the pairing Hamiltonian}
\author{T. Papenbrock}
\affiliation{Department of Physics and Astronomy, University of Tennessee,
Knoxville, TN~37996, USA}
\affiliation{Physics Division,
Oak Ridge National Laboratory, Oak Ridge, TN 37831, USA}
\author{Anirban Bhattacharyya}
\affiliation{Department of Physics and Astronomy, University of Tennessee,
Knoxville, TN~37996, USA}
\date{\today}

\begin{abstract}
We consider the pairing Hamiltonian and systematically construct its
density functional in the strong-coupling limit and in the limit of
large particle number. In the former limit, the functional is an
expansion into central moments of occupation numbers. In the latter limit,
the functional is known from BCS theory. Both functionals are nonlocal
in structure, and the nonlocalities are in the form of simple
products of local functionals. We also derive the relation between 
the occupation numbers and the Kohn-Sham density. 
\end{abstract}

\maketitle

\section{Introduction}
\label{intro}

Density-functional theory (DFT) is a very popular
theoretical method in many branches of physics, as it allows one to
obtain the ground-state energy of an interacting many-fermion system
from the solution of a mean-field equation~\cite{Hoh64,Koh65}.  The main
challenge in DFT is the connection between the Hamiltonian and the
form of the density functional. Usually, one approximates the unknown
(and supposedly nonlocal) density functional in terms of local
densities, currents, and gradients of the density. This approach has 
led to impressive results in quantum chemistry  \cite{Tao03}  
and nuclear physics \cite{Gor02,Ben03,Lun03,Sto03}. 

Pairing plays an important role in atomic nuclei (see, e.g.,
Ref.~\cite{Dean03} for a recent review). Within the mean-field approach
to nuclear structure, pairing is often included in the form of the BCS
model added to the Skyrme functional \cite{Vau73}, or through
Hartree-Fock-Bogoliubov (HFB) calculations~\cite{Dob84,Perl04}. The formal
extension of Kohn-Sham DFT to superconductors also leads to HFB
equations~\cite{Oli88}.  In these latter two approaches, the energy is
a functional of the density and the pairing field (anomalous
density). It is thought that the latter is a highly nonlocal
functional of the density and thereby captures aspects of the density
functional that would be difficult to model otherwise. Within the HFB
approach, the particle number is not a conserved quantity anymore, and
particle number projection becomes a concern and additional
expense~\cite{Ang01}.  In this work, we revisit the pairing problem
and construct its density functional without resorting to the HFB
approach.  We thereby avoid some of the problems associated with
pairing fields and gain insight into the nonlocal structure of the
density functional.

The direct and systematic construction of a density functional from a
given Hamiltonian is possible only for solvable or
sufficiently simple systems. We mention, for instance, the systematic
construction of the density functional for a dilute Fermi gas with
repulsive interactions ~\cite{Pug03,Bhat05}, or the description of the
Fermi gas close to the unitary limit in terms of local
densities~\cite{Car03} and gradient
corrections~\cite{Pap05,Bhat06}. Albeit being simple, these systems
are nontrivial, and they provide us with useful insights for the
empirical construction of the nuclear density functional. Examples
are, e.g., the occurrence of terms proportional to
$\rho^{7/3}$~\cite{Pug03} or proportional to $\rho^{5/3}$ \cite{Pap05}
which are absent in presently employed functionals.  The pairing
Hamiltonian is another solvable problem, and the derivation of its
density functional is the purpose of this paper.

This paper is organized as follows. In Sect.~\ref{meth}, we describe
the pairing Hamiltonian and the method to derive its density
functional.  In Sect.~\ref{bcs}, we construct the density functional in
the limit of large numbers of pairs.  In Sect.~\ref{G}, we construct
the density functional in the limit of large coupling strength. Both
density functionals will be given in terms of occupation numbers. In
Sect.~\ref{relate}, we work out the relation between occupation numbers
and the spatial density, and discuss extensions of presently employed
density functionals to include pairing. We close with a summary of our
results.

\section{Hamiltonian and method}
\label{meth}

The pairing Hamiltonian is defined as
\be
\label{ham}
H=\sum_{j=1}^\Omega\sum_{s=\downarrow,\uparrow} \varepsilon_j 
\hat{a}^\dagger_{js} \hat{a}_{js} - g \sum_{i,j=1}^{\Omega}
\hat{a}^\dagger_{i\downarrow} \hat{a}^\dagger_{i\uparrow} 
\hat{a}_{j\uparrow} \hat{a}_{j\downarrow}
\ .
\ee
We thus have $\Omega$ single-particle orbitals which can be doubly 
occupied with a constant pairing interaction. This
pairing model was studied and solved by Richardson in the 1960s as an
alternative to the BCS approximation \cite{Rich}. Recently, the
Richardson model has found renewed interest in the framework of
superconductivity in small mesoscopic systems, and we refer the reader
to the review by Dukelsky {\it et al.}~\cite{Duk04}. The Hamiltonian
(\ref{ham}) is exactly solvable and integrable (i.e., there are as many
conserved quantities as degrees of freedom~\cite{Sar97}). The ground-state 
energy  
\be
\label{E_rich}
E=2\sum_{i=1}^N E_i 
\ee
of $N$ pairs of fermions is given in terms of the 
solutions $E_i$ of the Richardson equations 
\be
\label{richeq}
\sum_{j=1\,(j\ne i)}^N {2\over E_i-E_j} = {2\over g} 
+ \sum_{k=1}^\Omega {1\over E_i-\varepsilon_k}\quad i=1,\ldots,N  \ .
\ee
Note that the ground-state energy~(\ref{E_rich})
depends implicitly on the single-particle energies $\varepsilon_k$.
Note also that the ground state of the
pairing model is a superposition of fully paired states, since
singly occupied orbitals are not subject to the interaction.

We want to construct the density functional for the pairing Hamiltonian. 
This requires us to compute the Legendre transform 
\be
\label{legendre}
F(\{n_k\}) = E - \sum_{k=1}^\Omega n_k \varepsilon_k
\ee
of the ground-state
energy (\ref{E_rich}) with respect to the occupation numbers
\be
\label{nocc}
n_k\equiv {\partial E\over \partial \varepsilon_k} \ .
\ee
Due to the Hellman-Feynman theorem, the occupation numbers (\ref{nocc}) 
are indeed the expectation value $n_k = 
\langle\psi| (\hat{a}^\dagger_{k\downarrow} \hat{a}_{k\downarrow} 
+\hat{a}^\dagger_{k\uparrow} \hat{a}_{k\uparrow})|\psi\rangle$
of the ground-state $|\psi\rangle$.  
Usually, DFT practitioners work with the density
instead of occupation numbers. However, 
the Hamiltonian~(\ref{ham}) is given in terms of the Fock-space
operators $\hat{a}^\dagger_k$ and $\hat{a}_k$, and its density functional
is therefore naturally expressed in terms of occupation. 
The relation between the density and the occupation
numbers will be given in Sect.~\ref{relate}. 

In order to actually perform the Legendre transform~(\ref{legendre}),
we need a closed expression of the ground-state energy in terms of the
single-particle energies. Unfortunately, no such expression is
available for the solutions of the Richardson
equations~(\ref{richeq}), as these equations have to be solved
numerically~\cite{Rom04,Dom06}.  However, closed expressions for the
ground-state energy do exist for the two limiting cases of a large
number of pairs $N\gg 1$ or a large coupling strength
$g/\varepsilon_k\gg 1$. In Sect.~\ref{bcs} and Sect.~\ref{G}, we will
construct the density functional of the pairing Hamiltonian in these
two limits, respectively.

Note that the density functional is only determined up to an overall
constant.  Shifting the single-particle energies $\varepsilon_k$ by a
constant does not change the corresponding occupation numbers
$n_k$. Indeed, summing Eq.~(\ref{nocc}) over $k$ yields $\sum_{k=1}^\Omega
n_k=2N$, which is independent of the single-particle energies
$\varepsilon_k$.  Thus, it is not possible to invert the
equations~(\ref{nocc}) and to express the single-particle energies in
terms of the occupation numbers. To avoid this problem, one can
introduce an orthogonal transformation from the single-particle
energies $\varepsilon_k$ to new variables $e_1,\ldots,e_{\Omega}$ such
that $e_1 = (1/\sqrt{\Omega})\sum_{k=1}^\Omega \varepsilon_k$. The new
variables $e_2,\ldots,e_\Omega$ are invariant under a constant shift of the
single-particle energies $\varepsilon_k$ (since they are orthogonal to
$e_1$).  The corresponding new occupation numbers $\nu_k=\partial
E/\partial e_k$ are related to the occupation numbers $n_k$ by the
same orthogonal transformation. One has in particular $\nu_1 =
2N/\sqrt{\Omega}$ and $F(\{\nu_k\})=E-\sum_k\nu_k e_k$.  The inversion
can be performed for the new variables $e_2,\ldots,e_{\Omega}$. Note
that variations of the resulting density functional $F(\{\nu_k\})$
with respect to $\nu_2,\ldots,\nu_{\Omega}$ conserve the number
of particles.

For the cases considered in Sect.~\ref{bcs} and in Sect.~\ref{G},
respectively, we can follow a simpler path. The ground-state energy
depends only on certain combinations of single-particle
energies. These can directly be expressed in terms of combinations of
occupation numbers without solving the individual $\varepsilon_k$
in terms of the $n_k$.

\section{Density functional in the limit of large particle number}
\label{bcs}

In the limit of a large number of pairs $N\gg 1$, the Richardson
equations~(\ref{richeq}) can be solved in a systematic expansion
\cite{Rich77,Yuz03}. This leads to the BCS energy
\be
\label{Ebcs}
E_{BCS} = \sum_{k=1}^\Omega \varepsilon_k -\mu (\Omega-2N) + {\Delta^2\over g} 
- \sum_{k=1}^\Omega\sqrt{(\varepsilon_k-\mu)^2 +\Delta^2} \ .
\ee
The gap $\Delta$ and the chemical potential $\mu$ determined by
\ba
\label{gap}
{2\over g} &=& \sum_{k=1}^\Omega 
{1\over \sqrt{(\varepsilon_k-\mu)^2+\Delta^2}} \ ,\\
\label{mu}
\Omega-2N &=& \sum_{k=1}^\Omega {\varepsilon_k-\mu 
\over \sqrt{(\varepsilon_k-\mu)^2+\Delta^2}} \ .
\ea

We are interested in the construction of the density functional for
the pairing Hamiltonian with ground-state energy~(\ref{Ebcs}) and
perform the Legendre transform~(\ref{legendre}) in three steps.  First,
we compute the occupation numbers
\be
\label{nbcs}
n_j={\partial E_{BCS}\over\partial\varepsilon_j} = 
1- {\varepsilon_j-\mu \over \sqrt{(\varepsilon_j-\mu)^2+\Delta^2}} \ .
\ee 
Here, we made use of Eqs.~(\ref{mu}) and (\ref{gap}). The form
Eq.~(\ref{nbcs}) is intuitively clear: For vanishing gap, the occupation
numbers are two and zero for orbitals below and above the chemical
potential, respectively. Second, we invert this equation and obtain the
single-particle energies as a function of the occupation numbers
\be
\varepsilon_j = \mu + {(1-n_j)\Delta\over \sqrt{n_j(2-n_j)}} \ .
\ee
Here, the gap and the chemical potential are also functions of the
occupation numbers. Indeed, summing Eq.~(\ref{nbcs}) over all orbitals
yields Eq.~(\ref{mu}), while we obtain 
\be
\Delta = {g\over 2} \sum_{j=1}^\Omega \sqrt{n_j(2-n_j)}
\ee 
from the gap equation~(\ref{gap}). Third, we compute the 
Legendre transform and obtain the density functional 
\ba
\label{df_bcs}
F_{BCS} &\equiv& E_{BCS} - \sum_{j=1}^\Omega \varepsilon_j n_j \nonumber\\
&=& -{\Delta^2\over g}\nonumber\\
&=& -{g\over 4} \left(\sum_{j=1}^\Omega \sqrt{n_j(2-n_j)}\right)^2 \ .  
\ea
The equations presented in this section are well known from BCS
theory~\cite{Lane64,Vau73}.  The form of the functional clearly
exhibits two properties of the pairing interaction. (i) Pairing
creates an instability of the Fermi surface, and (ii) is nonanalytical
for weak couplings. Property (i) is included in the
functional~(\ref{df_bcs}) as there is a gain of energy associated with
deviations of the occupation numbers from two and zero. Property (ii)
is also fulfilled since the functional is nonanalytical for small values
of $2-n_k$ and $n_k$, respectively. Note that the functional is
nonlocal, as it is the square of a local functional. Note also that
the functional~(\ref{df_bcs}) is among the simplest ones that includes
nonlocality, overall quadratic scaling in occupation numbers
(motivated by a two-body interaction), and properties (i) and
(ii). From this point of view, the form of the functional could almost
have been guessed.

\section{Density functional in the strong-coupling limit}
\label{G}

Pairing in nuclei is often associated with the weak-coupling regime.
An exception are, e.g., nuclei with magic proton number 
where the neutrons partly fill a
single $j$-shell. For such nuclei, pairing is dominant due to the
degeneracy of the single-particle orbitals. We consider the
corresponding strong-coupling regime in this section and gain
additional insight in the nonlocal structure of density functional.

Within the strong-coupling regime, the ground-state energy can be
computed within the quasi-spin formalism~\cite{Dob01}. Here, we follow
approach by Yuzbashyan {\it et al.}~\cite{Yuz03}. In the
strong-coupling limit, the pairing term is the dominant part of the
Hamiltonian~(\ref{ham}), and one has $g \gg
|\varepsilon_k-\overline{\varepsilon}|$. Here, and in what follows, we
denote averages over single-particle orbitals by the overbar, e.g.
\be 
\overline{\varepsilon}\equiv {1\over\Omega}\sum_{k=1}^\Omega
\varepsilon_k \ .  
\ee 
In the strong-coupling limit, the energies $E_i$ fulfill $E_i\gg
\varepsilon_k$, and one can expand the Richardson
equations~(\ref{richeq}) in terms of the small parameters
$\varepsilon_k/E_i$.  This yields sums of the form $\sum_i E_i^{-p}$
which in turn are expanded into a series of inverse powers of the
coupling $g$. The corresponding expansion coefficients can be
determined through recursion relations. We formally extend the results
by Yuzbashyan {\it et al.}~\cite{Yuz03} and write the the ground-state
energy as
\be
\label{expand}
E(\{\varepsilon_j\}) = E^{(1)} + 2N\overline{\varepsilon} + g \sum_{j\ge 2}
\sum_{\lambda} {c_{\lambda} \over g^j} \prod _{i=1}^k
\overline{\left(\varepsilon -\overline{\varepsilon}\right)^{\lambda_i}}
\ .
\ee
Here, $N$ denotes the number of fermion pairs, and 
\be
E^{(1)} \equiv -N(\Omega-N+1)g 
\ee
is the leading contribution of order $g$ to the ground-state energy.
The sum over $\lambda$ in Eq.~(\ref{expand}) runs over all partitions
$\lambda=[\lambda_1,\lambda_2,\ldots,\lambda_k]$ of $j$ into integers
$\lambda_1\ge\lambda_2\ge\ldots\ge\lambda_k\ge 2$ such that
\be
\label{lambda}
j = \sum_{i=1}^k \lambda_i \ .
\ee
Let us analyze the ground-state energy~(\ref{expand}). The first term
is of order $g$ and is quadratic in particle number. In the strong
coupling limit, each single-particle orbital has approximately equal
occupation $2N/\Omega$, and the second term on the
right hand side of Eq.~(\ref{expand}) sums the corresponding single-particle
energies. The subsequent terms contain ratios of single-particle
energies and the pairing strength, and these central moments are
invariant under shifts of the single-particle energies.

Yuzbashyan {\it et al.} estimated that the expansion (\ref{expand})
converges for ratios $(\varepsilon_j- \bareps)/ g\approx 1$. 
They gave explicit expressions for the first two coefficients
$c_{\lambda}$. Snyman and Geyer \cite{Sny06} corrected
a typo in the second coefficient, and Barbaro {\it et al.} \cite{Bar06}
gave an expression for the two fourth-order terms. We confirmed these
results by following the steps described in Ref.~\cite{Yuz03}. The first few
coefficients are thus
\ba
c_{[2]} &=& -{4N(\Omega-N)\over \Omega(\Omega-1)}\ ,\\
c_{[3]} &=& {8N(\Omega-N)(\Omega-2N)\over \Omega^2(\Omega-1)(\Omega-2)}\ ,\\
c_{[4]} &=& -{16N(\Omega-N)\over \Omega^3(\Omega-1)^2(\Omega-2)(\Omega-3)}
\nonumber\\
&&\times
\left( \Omega^2(\Omega-1) - N(\Omega-N)(5\Omega-6) \right)\ ,\\
c_{[2,2]} &=& {16N(\Omega-N)\over \Omega^3(\Omega-1)^2(\Omega-2)(\Omega-3)}
\bigg[\Omega^2(2\Omega-3)\nonumber\\
&&-N(\Omega-N)\left(3(3\Omega-4)-{\Omega\over
\Omega-1}\right)  \bigg]
\ ,
\ea
and higher-order coefficients can be worked out by following
the recursion relations given in Ref.~\cite{Yuz03}.

We are interested in the construction of the density 
functional~(\ref{legendre})
for the pairing Hamiltonian with ground-state energy $E$. This
requires us to perform a Legendre transform of the ground-state energy
(\ref{expand}) with respect to the single-particle energies (i.e. the
external potential). This transformation is done in three steps. 
First, we compute the occupation numbers
\be
n_j \equiv {\partial E(\{\varepsilon_k\})\over \partial \varepsilon_j} \ .
\ee
This is most easily done by utilizing the identity
\be
{\partial \over\partial \varepsilon_j} 
\overline{\left(\varepsilon -\overline{\varepsilon}\right)^k} =
{k\over\Omega}\left((\varepsilon_j -\overline{\varepsilon})^{k-1} -
\overline{\left(\varepsilon -\overline{\varepsilon}\right)^{k-1}}\right) \ .
\ee
Second, we express the single-particle energies as functions of the
occupation numbers. This inversion is most conveniently done order by
order in terms of the strong coupling expansion. Third, we have 
to compute the sum on the right hand side of Eq.~(\ref{legendre}). 
To this purpose we employ the identity
\be
\sum_{j=1}^\Omega n_j\varepsilon_j  =  
\sum_{j=1}^\Omega (n_j-\overline{n})(\varepsilon_j-\overline{\varepsilon})
+2N\overline{\varepsilon} \ ,
\ee
where
\be
\overline{n}\equiv {2N\over\Omega} 
\ee
is the average occupation number. 

Let us follow these steps order by order in the strong coupling expansion.
In leading order (LO), the ground-energy is
\be
\label{e_lo}
E_{\rm LO}(\{\varepsilon_j\}) = E^{(1)} + 2N\overline{\varepsilon}\ , 
\ee
and the occupation numbers~(\ref{nocc}) are
\be
n_j = \overline{n} \ .
\ee
This expression is independent of the single-particle energies, 
and the Legendre transformation cannot be performed. The formal reason is, 
of course, that the energy~(\ref{e_lo}) does only depend on the sum of the 
single-particle energies (cf. the discussion at the end of Sect.~\ref{meth}).
In the strong-coupling limit, all single-particle
orbitals have equal occupation. Thus, the individual occupation numbers do 
not depend on the single-particle energies.

In next-to-leading order (NLO), the ground-state energy is
\be
\label{e_nlo}
E_{\rm NLO} (\{\varepsilon_j\}) = E_{\rm LO}(\{\varepsilon_j\}) 
+ {c_{[2]}\over g}
\overline{\left(\varepsilon -\overline{\varepsilon}\right)^2} \ .
\ee
The occupation numbers fulfill
\be
\label{n_nlo}
n_j-\overline{n} = {2 c_{[2]} \over g\Omega}
(\varepsilon_j -\overline{\varepsilon}) \ .
\ee
Note that $c_{[2]}<0$, i.e. the occupation of orbitals below (above)
the average single-particle energy is above (below) the average occupation
number. For the inversion, we take the square of Eq.~(\ref{n_nlo}) and 
average over the single-particle orbitals. This yields
\be
\label{e2_nlo}
\overline{\left(\varepsilon -\overline{\varepsilon}\right)^2} = 
{g^2\Omega^2 \over 4 c_{[2]}}\overline{\left(n -\overline{n}\right)^2}\ .
\ee
Finally, we compute the sum
\be
\sum_{j=1}^\Omega n_j\varepsilon_j  = {2c_{[2]}\over g}
\overline{\left(\varepsilon -\overline{\varepsilon}\right)^2}
+2N\bareps \ .
\ee
Putting all together, we arrive at the density functional in NLO
\ba
\label{f_nlo}
F_{\rm NLO} (\{n_k\})&=& E_{\rm NLO} -\sum_{k=1}^\Omega n_k 
\varepsilon_k\nonumber\\
&=& E^{(1)} -{g\Omega^2\over 4c_{[2]}}
\overline{\left(n -\overline{n}\right)^2} \ .
\ea
In the presence of an external potential, given in terms of single-particle
energies $\varepsilon_j$, the density functional thus becomes
\be
E_{\rm NLO}(\{n_k\}) = F_{\rm NLO}(\{n_k\}) 
+ \sum_{j=1}^\Omega n_j\varepsilon_j \ .
\ee
The ground-state energy for the system with $N$ pairs 
is found from the requirement that
number-conserving variations of the functional vanish.
The corresponding occupation numbers fulfill
\be
{\partial E_{\rm NLO}(\{n_k\}) \over\partial n_j} = 0 \ ,
\ee
and it is understood that the derivative does not act on the $N$-dependent 
constants $E^{(1)}$ and $c_{[2]}$ of the functional~(\ref{f_nlo}).  
It is straightforward to check that the solution of this equation yields
the energy (\ref{e_nlo}).

For the calculation of the density functional in N$^2$LO we start from
the expression
\be
E_{\rm N^2LO} (\{\varepsilon_j\}) = E_{\rm NLO}(\{\varepsilon_j\}) 
+ {c_{[3]}\over g^2}
\overline{\left(\varepsilon -\overline{\varepsilon}\right)^3} 
\ee
for the energy, and repeat the three steps outlined above. For the
inversion, we raise the expression corresponding to Eq.~(\ref{n_nlo})
to the second and third power, and average over single-particle
orbitals. We neglect any terms of order
$O\left([(\varepsilon_j-\overline{\varepsilon})/g]^4\right)$ or higher 
that are generated in
this procedure.  This yields two linear equations that express the
central moments of occupation numbers $\overline{\left(n
-\overline{n}\right)^\mu}$ for $\mu=2,3$ in terms of the central moments of the
energy $\overline{\left(\varepsilon
-\overline{\varepsilon}\right)^\mu}$.  The inversion is straightforward, 
and the final result is
\be
\label{f_n2lo}
F_{\rm N^2LO} (\{n_k\})= F_{\rm NLO} (\{n_k\}) 
+ {g c_{[3]}\Omega^3\over 8 c_{[2]}^3}
\overline{\left(n -\overline{n}\right)^3} \ .
\ee 
Let us pause for a moment and consider Eqs.~(\ref{f_nlo}) and
(\ref{f_n2lo}), respectively.  The strong-coupling expansion of the
ground-state energy (\ref{expand}), given in central moments of the
single-particle energies, translates into an expansion of central moments of
occupation numbers. Clearly, deviations of the occupation numbers from
their mean value $\overline{n}$ increase with increasing value of the
ratios $(\varepsilon_j-\overline{\varepsilon})/g$. In the limit of
vanishing pairing strength $g\to 0$, the occupation numbers become
discontinuous at the Fermi energy. Note that the density functional is
local in the occupation numbers up to N$^2$LO, as only simple sums
over occupation numbers appear. Nonlocalities start to appear at
N$^3$LO, and we will address this order in what follows.

At N$^3$LO, the ground-state energy is 
\ba
E_{\rm N^3LO} (\{\varepsilon_j\}) &=& 
E_{\rm N^2LO}(\{\varepsilon_j\}) + {c_{[4]}\over g^3}
\overline{\left(\varepsilon -\overline{\varepsilon}\right)^4}\nonumber\\ 
&+& {c_{[2,2]}\over g^3}
\left(\overline{\left(\varepsilon -\overline{\varepsilon}\right)^2}
\right)^2 \ . 
\ea
As we will see, the last term in this equation
gives rise to the nonlocality in the density functional. Such
products of central moments of single-particle energies are ubiquitous at
higher orders.  For the occupation numbers, we find
\ba
n_j-\overline{n} &=& {2c_{[2]}\over g\Omega} 
(\varepsilon_j -\overline{\varepsilon})\nonumber\\&+& {3c_{[3]}\over g^2\Omega}
\left((\varepsilon_j -\overline{\varepsilon})^{2} -
\overline{\left(\varepsilon -\overline{\varepsilon}\right)^{2}}
\right)\nonumber\\
&+& {4c_{[4]}\over g^3\Omega}
\left((\varepsilon_j -\overline{\varepsilon})^{3} -
\overline{\left(\varepsilon -\overline{\varepsilon}\right)^{3}}\right)
\nonumber\\
&+&{4c_{[2,2]}\over g^3\Omega}
(\varepsilon_j -\overline{\varepsilon})
\overline{\left(\varepsilon -\overline{\varepsilon}\right)^{2}} \ .
\ea
We raise this equation to the power two, three, and four, average over
single-particle orbitals, and neglect any terms on the right hand
sides that are of order 
$O\left([(\varepsilon_j-\overline{\varepsilon})/g]^5\right)$ or
higher. The resulting three equations express the central moments of the
occupation numbers as nonlinear functions of central moments of the
single-particle energies, and the nonlinearities are due to the terms
proportional to $c_{[2,2]}$. We invert these equations and neglect
terms that are of order $O\left((n_j-\overline{n})^5\right)$ or
higher. The final result for the density functional is
\ba
\label{f_n3lo}
F_{\rm N^3LO} (\{n_k\})&=& F_{\rm N^2LO} (\{n_k\}) \nonumber\\
&+& {4c_{[2]}c_{[2,2]}+9c^2_{[3]}\over 64 c^5_{[2]}}g \Omega^4
\left(\overline{\left(n -\overline{n}\right)^2}\right)^2 \nonumber\\
&+& {4c_{[2]}c_{[4]}-9c^2_{[3]}\over 64 c^5_{[2]}}g \Omega^4
\overline{\left(n -\overline{n}\right)^4} \ .
\ea
The second term on the right hand side of this equation is nonlocal in
the occupation numbers, since it involves a double sum. This form of
nonlocality is, however, rather simple, as the double sum is a product
of two individual sums.

We can now generalize the expansion of the density functional to
higher orders.  The density functional at order $j$ consists of
products of central moments of the occupation numbers, and includes
occupation numbers up to the power of $j$, i.e
\be
\label{df_G}
F(\{n_j\}) = E^{(1)} + g \sum_{j\ge 2}
\sum_{\lambda} d_{\lambda}\Omega^j \prod _{i=1}^k
\overline{\left(n -\overline{n}\right)^{\lambda_i}}
\ .
\ee
Here, $\lambda$ again denotes the sum over partitions
(cf. Eq.~(\ref{lambda})), and coefficients $d_\lambda$ are functions
of the coefficients $c_\lambda$ in Eq.~(\ref{expand}). The functional
is nonlocal, but the multiple sums over occupation numbers
simply factor into products of simple sums. Thus, the nonlocal structure
is rather simple. It is also inexpensive from a computational point of view.

It interesting to compare the structure of the
functional~(\ref{df_G}) in the strong-coupling limit with the 
functional (\ref{df_bcs}) from BCS theory.  Within the BCS approximation,
the nonlocality is limited to the square of a local
functional. This suggests that products of multiple local functionals,
as they appear in (\ref{df_G}), must be corrections of order $1/N$ or
smaller. This can be verified by expanding the BCS
functional~(\ref{df_bcs}) around the strong coupling limit. We thus
set $n_j = \overline{n} + (n_j-\overline{n})$ in Eq.~(\ref{df_bcs})
and keep only terms up to second order in $n_j-\overline{n}$. The
result agrees with the NLO result~(\ref{f_nlo}) up to a term $-gN$.
This term is recovered when corrections of order $1/N$ to the
energy~(\ref{Ebcs}) are included (see, e.g. the first term of
Eq.~(2.16) in Ref.~\cite{Yuz05}).

\section{From occupation numbers to the density}
\label{relate}

How can one extend presently employed density functionals~\cite{Sk56,Vau72} 
to include
pairing effects? A number of approaches can be found in the
literature~\cite{Vau73,PG97,Taj04}. Here we follow the direct path
that links the occupation numbers to the density. Recall that 
DFT is built on the ground-state density
\be
\label{gsdens}
\rho(x)=\sum_{s=\downarrow,\uparrow}\langle\psi|\hat{\Psi}^\dagger_s(x) 
\hat{\Psi}_s(x)|\psi\rangle \ .
\ee
Here, $\hat{\Psi}_s^\dagger(x)$ creates a fermion with spin projection $s$ 
at the position $x$, and 
$|\psi\rangle$ denotes the ground state. However, the  
energy functionals~(\ref{df_bcs})
and (\ref{df_G}) are based on the occupation numbers $n_k$ that 
specify the occupation of the single-particle
orbitals of the pairing Hamiltonian (\ref{ham}). We denote the
corresponding single-particle wave functions as $u_k(x)$ and have 
\be
\hat{a}^\dagger_{ks} = \int d^3 x \,\,u_k(x)\hat{\Psi}^\dagger_s(x) \ .
\ee

Within Kohn-Sham DFT, the ground-state density~(\ref{gsdens}) 
is given in terms of the occupation numbers $n_\alpha$ of the
Kohn-Sham orbitals $\phi_\alpha(x)$, i.e. 
\be
\label{dens} 
\rho(x) = \sum_{\alpha} n_\alpha \phi^*_\alpha(x)\phi_\alpha (x) \ .  
\ee 
Here, $n_\alpha=2$ ($n_\alpha=0$) for occupied (unoccupied) Kohn-Sham
orbitals in fully paired systems, and we use the convention that Greek
indices denote Kohn-Sham orbitals while Roman indices label the
single-particle orbitals of the pairing Hamiltonian~(\ref{ham}).  The
occupation numbers $n_k$ are related to the Kohn-Sham orbitals and
Kohn-Sham occupation numbers via
\be
\label{n_ks}
n_k = \sum_\alpha \langle k|\alpha\rangle n_\alpha \langle \alpha|k\rangle \ , 
\ee
with
\be
\label{overlap}
\langle \alpha |k\rangle = \int d^3x \phi^*_\alpha(x) u_k(x) \ .
\ee
Thus, the occupation numbers (\ref{n_ks}) are functionals of the 
Kohn-Sham orbitals and therefore nonlocal functionals of the density. 
The functional derivative of interest is
\ba
\label{deriv}
{\delta F(\{n_k\})\over \delta \phi^*_\alpha(x)} &=& \sum_k 
{\partial F\over\partial n_k} 
{\delta n_k\over \delta \phi^*_\alpha(x)} 
\nonumber\\
&=& n_\alpha \sum_k  {\partial F\over\partial n_k} 
\langle k|\alpha\rangle u_k(x) \ .
\ea
Thus, one might add a functional of the form~(\ref{df_G}) or
(\ref{df_bcs}) to commonly employed density functionals~\cite{Sk56,Vau72}. The
single-particle wave functions $u_k(x)$ are determined by the pairing
Hamiltonian; they can be, e.g., plane waves or shell model
orbitals. The nonlocality of the density functional has two
sources. The first is the nonlocal structure of the
functionals~(\ref{df_G}) and (\ref{df_bcs}) in terms of the occupation
numbers, and the second is due to the relation~(\ref{n_ks}) between
the occupation numbers and the Kohn-Sham orbitals.

The use of occupation numbers~(\ref{n_ks}) in a density functional is
only new at first sight. The Kohn-Sham kinetic energy
\ba
\label{kin}
T &=& 
{\hbar^2\over 2m} \sum_\alpha n_\alpha \int d^3 x \nabla \phi^*_\alpha(x) 
\cdot \nabla \phi_\alpha(x) \nonumber\\
&=& \sum_k {\hbar^2 k^2\over 2m} n_k
\ea
can also be written in terms of occupation numbers $n_k$ of plane wave
states.  Equation~(\ref{kin}) is obtained through integration by parts
and by a Fourier expansion of the Kohn-Sham orbitals $\phi_\alpha(x)$
in terms of plane waves $u_k(x)$. Thus, DFT practitioners have dealt
with this form of nonlocality for a long time. This insight opens the
avenue for more complex (and hopefully more precise) density
functionals for nuclei. Recall that the very accurate mass formula by
Duflo and Zuker \cite{Duf95} explicitly depends on the number of
valence nucleons outside of closed shells. These are occupation
numbers of shell-model orbitals, e.g., orbitals $u_k(x)$ from the
harmonic oscillator or the Woods-Saxon Hamiltonian. 
Equation~(\ref{n_ks}) might permit us to explore and extend this very
successful mass formula in the DFT setting.

The main advantage of the functionals~(\ref{df_bcs}) and (\ref{df_G})
over the HFB approach consists of the conservation
of particle number. This is very encouraging, as number projection
techniques are numerically expensive and pose practical and
theoretical difficulties \cite{Dob05}. However, the
Hamiltonian~(\ref{ham}) is not realistic, and a constant pairing
interaction is a crude approximation for nuclei. Furthermore, the
involved sums over single-particle orbitals do not converge for
infinitely large model spaces. To avoid divergences, one might
introduce a cutoff (e.g., the Debye frequency for condensed matter
problems), or utilize regularization and renormalization procedures
that relate the coupling $g$ to physical observables
\cite{Stri98,Pap99,Bul02,Dug04}.  These methods not only cure
divergences, they also suggest how to make the density
functionals~(\ref{df_bcs}) and (\ref{df_G}) more realistic. The
density functionals of the pairing Hamiltonian are totally symmetric
in the occupation numbers $n_k$. This symmetry, of course, reflects
the fact that the pairing interaction~(\ref{ham}) is a constant and
equally scatters pairs between the single-particle orbitals. This
suggests to modify the functional~(\ref{df_bcs}) through
\be
\label{df_modify}
\sum_k \sqrt{n_k(2-n_k)} \to 
\sum_k w_k \sqrt{n_k(2-n_k)}  \ .  
\ee
Here, $w_k$ is an orbital-dependent weight or cutoff function. The 
modified functional still exhibits the hallmarks of pairing,
as they are described below Eq.~(\ref{df_bcs}); appropriately chosen 
weight functions $w_k$ also regularize the sums over
single-particle orbitals, and should be capable
of describing more realistic situations.

\section{Summary}

We constructed the density functional of the pairing Hamiltonian. This
construction is possible in the strong coupling limit, and within the
BCS approximation. In the strong coupling limit, the functional is a
multiple product of local functionals that depend on the central
moments of the occupation numbers. In the BCS
approximation, the functional is arguably the simplest nonlocal and
nonanalytical functional that causes an instability of the Fermi
surface. Both functionals are based on occupation numbers of
single-particle orbitals that define the pairing Hamiltonian. These
occupation numbers are themselves functionals of the Kohn-Sham
orbitals that define the density. Our results suggest a way to
include pairing with particle-number conservation into nuclear
density-functional theory.

We thank W. Nazarewicz and M. Ploszajczak for useful discussions.
This research was supported in part by the U.S. Department of Energy
under Contract Nos.\ DE-FG02-96ER40963 (University of Tennessee) and
DE-AC05-00OR22725 with UT-Battelle, LLC (Oak Ridge National
Laboratory).

\end{document}